\documentclass[%
reprint,
 amsmath,amssymb,
 aps,
prb,
]{revtex4-2}

\usepackage{graphicx}
\usepackage{dcolumn}
\usepackage{bm}

\usepackage[T2A]{fontenc}
\usepackage[utf8]{inputenc}

\usepackage[usenames]{color}
\usepackage{colortbl}
\usepackage{amsmath}
\usepackage{amssymb}

\begin{document}

\preprint{APS/123-QED}

\title{Phase diagram of a ferromagnetic semiconductor. The origin of superparamagnetism.}

\author{N. A. Bogoslovskiy}
\author{P. V. Petrov}
 \email{pavel.petrov@gmail.com}
\author{N. S. Averkiev}
\affiliation{Ioffe Institute, Russian Academy of Science, 194021 St. Petersburg, Russia}

\date{\today}

\begin{abstract}

We study the theoretical model of a ferromagnetic semiconductor as a system of randomly distributed
Ising spins with a long-range exchange interaction. Using the density-of-states approach, we
analytically obtain the magnetic susceptibility and heat capacity over a wide range of temperatures
and magnetic fields. It is shown that the finite system of spins in magnetic field less than
a certain critical field is in a superparamagnetic state due to thermodynamic fluctuations.
The complex phase structure of a ferromagnetic semiconductor is discussed.

\end{abstract}

\maketitle
\section{Introduction}
During the past years, significant progress has been made in the field of ferromagnetic semiconductor materials. 
In the pioneering Ohno's work~\cite{10.1063/1.118061}, it was demonstrated that ferromagnetism in GaAs
doped with Mn is associated precisely with the properties 
of the doped semiconductor, and not with the presence of MnAs inclusions. 
Since that moment, the list of ferromagnetic semiconductor materials has been constantly expanding,
while the experimentally observed values of the Curie temperature 
have reached room temperature \cite{PhysRevLett.99.127201, PhysRevLett.117.227202}. 
On the other hand, a clear theoretical description of ferromagnetism in semiconductors has not yet been achieved. 
Various mechanisms of ferromagnetic exchange between impurities have been proposed: exchange mediated by delocalized holes \cite{PhysRevB.63.195205}, 
percolation of bound magnetic polarons \cite{PhysRevLett.88.247202} and hopping mechanism \cite{PhysRevLett.99.227205}.
But even for GaAs:Mn, the most studied to date, there is no theory covering all the main experimental observations. 
It is not even clear whether the valence or conduction band is responsible for ferromagnetism. \cite{10.1038/nmat3317}.

One of the most significant differences of semiconductor ferromagnetic
materials is the random distribution of interacting spins, while in
conventional magnetics they are located at the nodes of a regular lattice.
Spin systems with spacial disorder have already been studied theoretically, but mostly
with antiferromagnetic sign of interaction and by means of numerical simulation
\cite{PhysRevLett.48.344, McLenaghan_1984, doi:10.1063/1.335065,
bogoslovskiy2019impurity, bogoslovskiy2021spin}.

In a number of works \cite{OHNO1999110, sawicki2010experimental, PhysRevB.94.075205, PhysRevMaterials.1.054401, Yuan_2018, PhysRevB.97.115201, 10.1063/5.0031605}, 
it was experimentally shown that, a ferromagnetic transition in doped semiconductors has a complex nature. As the concentration
of magnetic impurities increases, a material first passes from the paramagnetic to the superparamagnetic, and then to the ferromagnetic phase. 
The most probable scenario of such a transition is follows. 
Due to the random distribution of magnetic impurities in the sample, there are regions with a local concentration 
higher than a certain critical concentration. Such regions we call clusters for brevity.
The exchange interaction aligns the spins inside the clusters in one direction.
Due to thermodynamic fluctuations, which could not be neglected in finite systems,
the magnetic moment of such clusters is not fixed, and they behave like superparamagnets.
As the concentration increases the growing clusters merge in one macroscopic ferromagnet.
The purpose of our work is to theoretically investigate the physical properties of these clusters depending on temperature,
magnetic field and the number of spins in the cluster.

Here we show that the statistical approach makes it possible
to analytically calculate the density of states $g(E, M)$ of the cluster of randomly distributed spins
as a function of the total exchange energy $E$ and magnetic moment $M$. It should be
emphasized that, in contrast to the one-electron density of states,
here we are talking about the states of the entire
system of spins, the total exchange energy, and the total magnetic moment
of the system.

If the density of states is known, it is easy to find the partition function and other physical properties of the system.
In this approach $g(E, M)$ with a given $M$ is considered as a probability density function
of the distribution of total exchange energy.
For the first time this method was introduced by Heisenberg in his pioneering
article about the nature of ferromagnetism~\cite{Heisenberg1928}. Later it has been applied in the spin glass researh~\cite{PhysRevB.24.2613}.
Recently similar approach has been used to study the Ising problem on a regular lattice in high dimensions~\cite{e23121665}.
The difference between our model and mentioned researches is that we take into account a structural disorder and consider large but finite system.
The fact that the moments of distribution depend on $M$ imply that disorder in our model depends on the temperature and magnetic field.
It means that we treat disorder as annealed in contrary with a spin glass approach which usually treat the disorder as quenched~\cite{PhysRevB.24.2613}.

\section{Model}
We consider a finite system of $N$ randomly distributed spins rigidly fixed in space. 
We use an Ising model with a ferromagnetic long-range interaction $J(r)$ to describe the energy of the system.
Each spin can be in one of two states with a magnetic moment $\mu s$ where $s=\pm1$. 
Then the Hamiltonian of the system is given by:
\begin{equation}
H = -\frac{1}{2} \sum_{i \ne j} J(r_{ij}) s_i s_j - \sum_{i} \mu B s_i 
\label{hamiltonian}
\end{equation}

The first term is the total exchange energy
\begin{equation}
E = -\frac{1}{2} \sum_{i \ne j} J(r_{ij}) s_i s_j = 
-\frac{1}{2} \sum_i J_i s_i,
\label{exchange_energy}
\end{equation}
where $- J_i s_i = - s_i \sum_{j \ne i} J(r_{ij})s_j$ is the exchange interaction energy of the spin $i$ with all other spins. 
All calculations here are performed with the hydrogen-like dependence of
the exchange energy $J$ on the distance \cite{1964Gorkov, PhysRev.134.A362} in a three-dimensional space.
\begin{equation}
J(r) = J_0 \left( \frac{r}{a} \right)^{5/2} \exp{\left(-\frac{2r}{a} \right)},
\label{hydrogen}
\end{equation}
where $a$ is the Bohr radius. However, the solution can be easily generalized to a wider class of functions $J(r)$ (RKKY-type, for instance),
as well as to an arbitrary finite space dimension.

The total exchange energy is the sum of $N$ random identically distributed energies $-J_i s_i$.
In accordance with the central limit theorem, the distribution of $E$ converges to the normal distribution as the number of spins in the system increases.
It is known that for the finite sum of non-gaussian random variables the distribution tails deviate from the normal one~\cite{PhysRevLett.89.070201}.
However, in one of the previous papers, we have shown numerically that if the spin concentration $na^3$ is higher
than a certain critical concentration $n_c a^3$, the one-spin energy $-J_i s_i$ has a normal distribution \cite{bogoslovskiy2021spin}. 
In accordance with Cram{\'e}r's theorem~\cite{Cramer1936-wy, cramer1999mathematical} the sum of normally distributed one-spin energies is also
normally distributed. In that case, the gaussian approximation is applicable. 
Analytically the critical concentration $n_c a^3$ could be estimated using the central limit theorem
in Lindeberg's formulation~\cite{Lindeberg1922-gr, cramer1999mathematical}.
The distribution of a sum of random variables is normal if none of the terms makes a dominant contribution to the sum.
The maximum contribution to the one-spin energy comes from the terms for which the value of $4\pi r^2 J(r)$ is maximal.
For $J(r)$ of the form (\ref{hydrogen}), this distance is $9a/4$.
If it is greater than the average distance to the nearest neighbor, 
then the contribution of the nearest neighbor is not dominant, and the one-spin energy distribution is normal.
Using the formula for average distance to the nearest neighbor $\overline{r} = \Gamma(4/3)/(4\pi n/3)^{1/3}$ from \cite{RevModPhys.15.1}, 
one can estimate the critical concentration $n_{c}a^3 = 0.015.$

In order to determine the density of states, it is necessary to derive
the average energy $\overline {E_m}$ and the variance $\sigma_m$ for each $m$ averaged over
random distribution of spins in space. Here we denote the number of “down” spins by $q$, the number of “up” spins by $N - q$, and
the dimensionless magnetic moment per one spin by $m =\frac{M}{\mu N} =1-\frac{2q}{N}$.
The system of $N$ spins in the Ising model has $2^N$ possible states, and
the number of states with a fixed value of the magnetic moment is equal to the binomial coefficient $\binom{N }{q}$.
If we assume that these states are normally distributed in energy, the density of states with a given $m$ is
\begin{equation}
g_m(E) = \binom{N}{\frac{N(1-m)}{2}} \frac {1} {\sqrt {2 \pi } \sigma_m} 
\exp \Biggl( {- \frac { \left ( E - \overline E_m \right ) ^2} {2 \sigma_m^2}} \Biggr)
\label{gauss}
\end{equation}
Averaging over configurations, one can replace the sum over discretely located spins
by an integral over space with a uniform distribution of the magnetic moment with a density $nm$.
The average energy in the limit of large $N$ is
\begin{gather}
\overline E_m = -\frac{N}{2} \overline {s_i} \overline {J_i} =
- \frac{mN}{2} \int \limits_0^\infty {n m J(r) 4\pi r^2 dr} =
- \frac{m^2 N}{2} \overline {J_{1}}  
\nonumber  \\
\overline {J_{1}} = \frac{945 \pi}{2^{8}} \sqrt {\frac {\pi}{2}} J_0 n a^3
\label{mean_E}
\end{gather}

Using the same line of reasoning, after cumbersome calculations which are described in the Appendix,
we obtain the following expression for the variance
\begin{gather} 
\sigma_m^2 = \overline {E^2} - \overline {E}^2 =
\left( 1 - m^4 \right) N \sigma^2_1 \nonumber \\
 \sigma^2_1 = \frac{7!}{2^{15}} \pi J_0^2 n a^3
\label{sigma}
\end{gather}

It is important that all further calculations do not depend on the form of $J(r)$. 
It is only necessary that the values $\overline {J_{1}}$ and $\sigma_1$ be finite and the concentration exceeds $n_c$.

Using the Stirling formula the binomial coefficient in (\ref{gauss}) can be rewritten as
\begin{equation}
\binom{N}{\frac{N(1-m)}{2}} = \sqrt{\frac{2}{\pi N}} \frac {1}{\sqrt{1 - m^2}} \exp \left( N p(m) \right)
\end{equation}
Here we introduce the notation
\begin{equation} 
\nonumber
p(m) = \ln 2 - \frac{1-m}{2} \ln(1-m) - \frac{1+m}{2} \ln(1+m)
\end{equation}

For convenience we introduce a dimensionless energy per one spin $e = \frac{E} {N \overline {J_1}}$
and a dimensionless standart deviation $s_1 = \frac{\sigma_1}{\overline {J_1}}$.  
For large $N$, we assume that the average magnetic moment $m$ varies continuously in the range from -1 to 1, 
and determine the density of states in terms of energy and magnetic moment.
\begin{equation}
\begin{aligned}
g(e, m) = \frac{N}{2 \pi s_1 (1 - m^2) \sqrt{1 + m^2}} \times  \\
\exp \left ( { N p(m) - \frac { N \left ( e + m^2 / 2 \right ) ^2} {2 s_1^2 (1-m^4)}} \right ) 
\label{gauss2}
\end{aligned}
\end{equation}
\begin{figure}[t]
\centering
\includegraphics[width=1\linewidth]{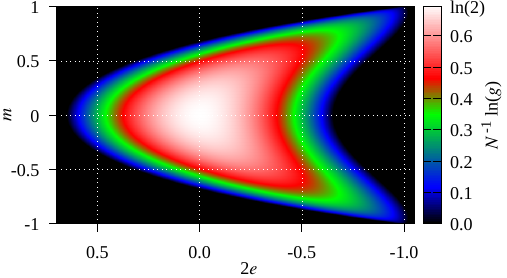}
\caption{Logarithm of the density of states $g(e, m)$ normalized to the number of spins $N$, $na^3=0.03$.}
\label{g_em}
\end{figure}
It is noteworthy that $N^{-1} \ln g(e, m)$ in the limit of large $N$ is universal and does not depend on $N$ (figure \ref{g_em}).

\section{Numerical simulation}

The formulas (\ref{gauss}--\ref{sigma}) were independently verified by two numerical methods, the Wang-Landau and the direct sampling method. 
The Wang-Landau algorithm \cite{PhysRevLett.86.2050, doi:10.1119/1.1707017} is a non-Markovian random walk in the phase space, taking into account the statistics of previous visits. 
The calculations were carried out using parallel computing, the density of states was calculated separately for the limited set of $m$ values~\cite{10.18721/JPM.161.301}. 
Random walks are performed by simultaneously flipping two randomly chosen antiparallel oriented spins in order to keep $m$ constant. 
Our calculations for $N=8192$ showed that  $g(e, m)$ for each $m$ has the form of a normal distribution with insignificant deviations on the distribution tails. 
The calculated dependences of the average energy and variance on $m$ agree well with the theoretical
values (\ref{mean_E}) and (\ref{sigma}).
\begin{figure}[t]
\centering
\includegraphics[width=1\linewidth]{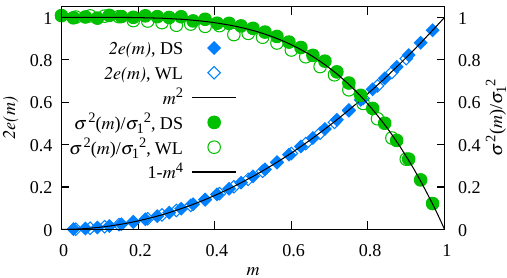}
\caption{Dependence of the mean exchange energy and the variance on the magnetic moment calculated by Wang-Landau (WL)
and direct sampling (DS) numerical methods in comparision with equations (\ref{mean_E}) and (\ref{sigma}).}
\label{fig:mpr}
\end{figure}

The direct sampling method consists in sequential calculation of the system energies with
a random spin configuration, but with a fixed value of $m$. 
After the accumulation of a sufficiently large number of samples, the first four central moments of the distribution were calculated using the obtained samples of energy. 
The total energy of the spin system (\ref{exchange_energy}) can be represented as $E(\mathcal{S}) = 1/2\; \mathcal{S}^{\rm T} \mathcal{J}\, \mathcal{S}$. 
Here $\mathcal{J}$ is the matrix of interaction energies $J_{i,j}$ of spins $i$ and $j$, $\mathcal{S}$ is the column of spin variables. 
Using parallel computing on the GPU with the implementation of CUDA technology for the julia language \cite{besard2018juliagpu},
we were able to significantly increase the performance of scalar product calculations
and increase the size of the system up to $N=32768$. 
Obtained dependences of average energy $\overline{E}(\mathcal{S})$ and variance $\sigma^2 = \overline {E^2}(\mathcal{S}) - \overline{E}(\mathcal{S})^2 $ also agree 
with the theoretical formulas (\ref{mean_E}, \ref{sigma}), while the third and fourth moments do not depend on $m$ and are equal to $0 \pm 0.01$ and $3 \pm 0.02$, respectively as expected for a normal distribution.
The combined results of numerical computations in comparision with theoretical equations (\ref{mean_E}) and (\ref{sigma})
are presented on figure~\ref{fig:mpr}.
Averaging over a spatial distribution has particular difficulty compared to averaging over other types of disorder.
Each realization of spin coordinates corresponds to a slightly different effective spin concentration.
For this reason, we study only one realization of structural disorder using both numerical methods.
The deviation of numerical simulation results from theoretical curves which can be seen on figure~\ref{fig:mpr}
is due to the fact that energies of single realization could not be considered as completely independent random variables.
The direct sampling method demonstrates better agreement with the theory due to larger $N$ and better self-averaging.

\section{Connection with Curie-Weiss theory and the Landau theory}

In what follows we consider a system with the density of states given by (\ref{gauss2}). 
Let the system have temperature $T$ and be in an external magnetic field $B$.
The probability for the system to be in a certain state can be described by the Boltzmann distribution with energy $E - \mu B N m$.
Here, as above, $E$ denotes only the exchange energy. 
For convenience, we introduce a dimensionless temperature $t = kT / \overline {J_1}$ and a dimensionless magnetic field $\beta = \mu B / \overline {J_1}$.
In this notation the probability density for the system to have energy $e$ and magnetic moment $m$ at temperature $t$ and in an external magnetic field $\beta$ is
\begin{equation}
\begin{aligned}
f (e, m, t, \beta) = \frac{1}{Z(t, \beta)} \frac{N}{2 \pi s_1 (1 - m^2) \sqrt{1 + m^2}} \times \\ 
 \exp \left ( { N p(m) - \frac { N \left ( e + m^2 / 2 \right ) ^2} {2 s_1^2 (1-m^4)}} - \frac {N \left( e - \beta m \right)} {t} \right ) 
\end{aligned}
\label{probability}
\end{equation}
Here $Z=\iint g(e, m) e^{-N(e-\beta m)/t} de\,dm$ is the partition function. After intergration over energy
\begin{equation}
\begin{aligned}
Z(t, \beta) = \sqrt{\frac{N}{2 \pi}}  \int \frac{1}{\sqrt{1 - m^2}}  \times \qquad \qquad \\
\exp {\left ( N p(m) + N\frac {s_1^2 (1 - m^4) + m^2 t + 2 \beta m t} {2 t^2} \right )} dm
\label{partition}
\end{aligned}
\end{equation}

Let us demonstrate the connection between our method and conventional approaches
such as the Curie-Weiss theory and the Landau theory.
In the case of large $N$ the integral over $m$ in (\ref{partition}) can be calculated analytically using the Laplace's method.
The value $m_0$ which correspond to the maximum of the exponent can be found from 
\begin{equation}
- \frac{2 s_1^2 m_0^3}{t^2} + \frac{m_0}{t} + \frac{\beta}{t} + \frac{1}{2} ln \left( \frac{1-m_0}{1+m_0} \right) = 0
\label{m0}
\end{equation}

Equation (\ref{m0}) can have one or three roots in depends on $t$ and $\beta$.
First we consider the case when the equation (\ref{m0}) has one root. 
The average magnetic moment calculated by the Laplace's method is $\overline{M} (t, \beta) = \mu N m_0$. 
The magnetic susceptibility can be obtained by dividing the variables and differentiating the equation (\ref{m0}).
\begin{equation} 
\chi = \frac{\partial \overline{M}}{\partial B}  = 
\frac{\mu^2 N}{\overline{J_1}}  \frac {(1-m_0^2)}{t - (1-m_0^2) + \frac{6 s_1^2}{t} (m_0^2 -m_0^4)}
\label{chi}
\end{equation}

In weak magnetic fields $\beta \ll 1$ the value of $m_0$ is small and the magnetic susceptibility (\ref{chi}) converges to
\begin{equation}
\chi = \frac{\mu^2 N}{\overline{J_1} (t - 1)} 
\end{equation}

Note that this expression coincides with the Curie-Weiss law with the Curie temperature $t_c = 1$.

After integration (\ref{partition}) using the Laplace's method
\begin{gather}
Z(t, \beta) = \sqrt {\frac{t^2}{ \left( 6 s_1^2 m_0^2 - t \right) (1 - m_0^2) + t^2}} \times \nonumber \\
\exp {\left ( N \left( \frac {s_1^2 (1-m_0^4) + m_0^2 t + 2 \beta m_0 t} {2 t^2} +  p(m_0) \right) \right)}
\label{Z0}
\end{gather}

In the limit of large $N$, the pre-exponential factor in the expression (\ref{Z0}) could be discarded.
In a weak magnetic field $m_0 \ll 1$ and $p(m_0) \approx \ln 2 - \frac{m_0^2}{2} - \frac{m_0^4}{12}$.
In these approximations, the thermodynamic free energy $F\,=\,-kT\,ln\,Z$ is
\begin{equation}
\begin{aligned}
F(t, \beta) =  N \overline {J_1} \biggl(- t \ln 2 - \frac {s_1^2}{2t} + \quad \\
+ \frac{t - 1}{2} m_0^2 + \left( \frac {s_1^2}{2t} + \frac{1}{12} \right) m_0^4  - \beta m_0 \biggr)
\label{Landau}
\end{aligned}
\end{equation}

This expression coincides with the Landau's theory of phase transitions \cite{landau2013statistical},
and the average magnetic moment $m_0$ has the meaning of the order parameter.
The phase transition from the paramagnetic to the ferromagnetic phase occurs at a temperature $t = 1$, when the coefficient of $m_0^2$ changes its sign.
It is noteworthy that our model is applicable for arbitrary values of the magnetic field, not only in the limit
of low fields as in the Landau theory.

\begin{figure}[h]
\includegraphics[width=1\linewidth]{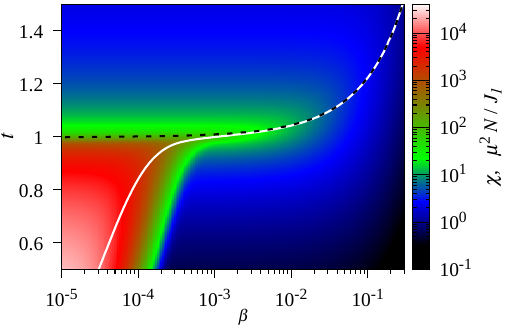}
\caption{The color map shows the magnetic susceptibility in coordinates ($t, \beta$). The white line shows the maximum magnetic susceptibility as a function of temperature at each value of the magnetic field, $N=16384, na^3=0.03$.}
\label{magnetic_susceptibility}
\end{figure}
\begin{figure}[h]
\includegraphics[width=1\linewidth]{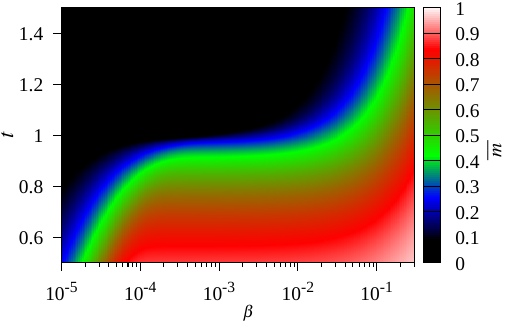}
\caption{The color map shows the average magnetic moment in coordinates ($t, \beta$), $N=16384, na^3=0.03$.}
\label{magnetic_moment}
\end{figure}

\section{Magnetic susceptibility and average magnetic moment}

If (\ref{m0}) has three roots, the exponent (\ref{partition}) has two local maxima. 
Let's denote the corresponding roots of the equation (\ref{m0}) as $m_+$ and $m_{-}$. 
In this case, the partition function can also be calculated using the Laplace's method similar to (\ref{Z0}). 
The partition function is expressed as the sum of two terms, which we denote as $Z_+$ and $Z_{-}$, respectively.
In this notation the average magnetic moment is
\begin{equation}
\overline{M} = \mu N \frac{m_+ Z_+ + m_- Z_-}{Z_+ + Z_-}
\label{mean_M}
\end{equation}

The Laplace's method is not applicable in the vicinity of the phase transition.
However, the average magnetic moment and susceptibility at an arbitrary temperature can be
calculated numerically. $$\overline m
(t, \beta) = \iint m f (e, m, t, \beta) de~dm; \;\;\;\;\;\; \chi = \frac{\mu^2 N}{J_1} \frac{\partial \overline{m}}{\partial \beta} $$

Figure \ref{magnetic_susceptibility} shows the magnetic susceptibility.
The white line is the maximum versus temperature at given magnetic field.
In high fields, the maximum shifts to higher temperatures and noticeably broadens.
In weak magnetic fields, the maximum shifts strongly down in temperature. 
This effect strongly depends on $N$ and drastically differs from the behaviour of the infinite system where the maximum
converges to $t = 1$ for small $\beta$ (black dashed line on figure \ref{magnetic_susceptibility}).

Figure \ref{magnetic_moment} shows the corresponding dependence of an average magnetic moment on magnetic field and
temperature. It is noteworthy that in small magnetic fields the average magnetic moment retains zero
at temperatures is much lower than the Curie temperature.
Nevertheless, fluctuations of magnetic moment in this region are significant.

\section{Magnetic moment fluctuations}

\begin{figure}[b]
\includegraphics[width=1\linewidth]{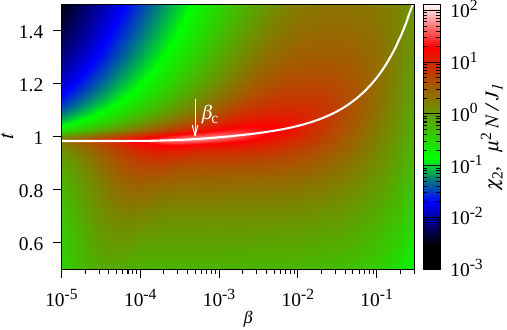}
\caption{The color map shows $\chi_2$ in coordinates ($t, \beta$).
The white line shows the maximum $\chi_2$ as a function of temperature at each magnetic field value.
The arrow shows a magnetic field $\beta_c$ which corresponds to the absolute maximum of $\chi_2$, $N=16384, na^3=0.03$.}
\label{chi_2}
\end{figure}
\begin{figure}[h]
\includegraphics[width=1\linewidth]{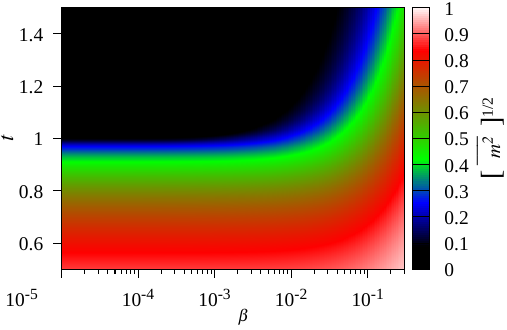}
\caption{The color map shows $\sqrt{\,\overline{m^2}}$ in coordinates ($t, \beta$). $N=16384, na^3=0.03$.}
\label{magnetic_moment_2}
\end{figure}
\begin{figure}[h]
\includegraphics[width=1\linewidth]{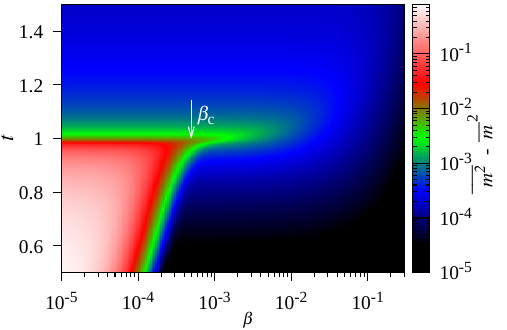}
\caption{The color map shows the variance of magnetic moment $\overline{m^2} - \overline{m}^2$ in coordinates ($t, \beta$).
The arrow shows a magnetic field $\beta_c$ which corresponds to the absolute maximum of $\chi_2$, $N=16384, na^3=0.03$.}
\label{fluctuations}
\end{figure}

In order to reveal the picture of magnetic moment fluctuations we use the special kind of susceptibility which we define as
$$\chi_2 = \frac{\mu^2 N}{J_1} \frac{\partial \sqrt{\,\overline{m^2}}}{\partial \beta} $$
Here we use $\sqrt{\,\overline{m^2}}$ as a fluctuation order parameter~\cite{demishev2020electron, bogoslovskiy2021spin, demishev2022spin}. 
Figures \ref{chi_2} and \ref{magnetic_moment_2} show the dependences of $\chi_2$ and $\sqrt{\,\overline{m^2}}$
on magnetic field and temperature. In contrast to $\overline{m^2}$ (see figure~\ref{magnetic_moment}) there is no region at $t<1$ in which
$\sqrt{\,\overline{m^2}}$ close to zero in any magnetic field. The dependence of $\chi_2$ has
a maximum at certain $\beta_c$ and $t_c$ which shifts towards the point $\beta=0$, $t=1$
while $N$ increases. We associate the region with the almost zero magnetic moment and strong magnetic fluctuations with superparamagnetic state of
the spin system. Figure \ref{fluctuations} which shows dependence of the magnetic moment variance $\overline{m^2} - \overline{m}^2$
illustrates this statement. There is an area with significant fluctuations of magnetic moment at small $\beta$ and $t$.

\section{Heat capacity}

In the same way as for magnetic susceptibility, we find the average exchange energy per spin and the heat capacity
by means of numerical integration.
$$ \overline e (t, \beta) = \iint e f (e, m, t, \beta) de~dm; \;\;\;\;\;\;\;\; C= kN \frac{\partial \overline e}{ \partial t} $$
\begin{figure}[t]
\includegraphics[width=1\linewidth]{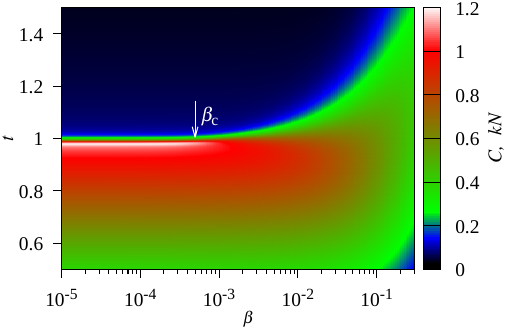}
\caption{The color map shows the heat capacity in coordinates ($t, \beta$).
The arrow shows a magnetic field $\beta_c$ which corresponds to the absolute maximum of $\chi_2$, $N=16384, na^3=0.03$.}
\label{heat}
\end{figure}

Figure \ref{heat} shows the heat capacity in coordinates ($t, \beta$).
It is noteworthy that, in contrast to the magnetic susceptibility, the maximum of the heat capacity is close to $t =1$ 
and constant in low magnetic fields. In the region of high values of $\chi_2$ the heat capacity has a narrow cusp which
gradually disappears in magnetic fields $\beta > \beta_c$.  At high fields, the maximum of the heat capacity
shifts towards high temperatures, as the maximum of magnetic susceptibility.

\section{Phase diagram}

\begin{figure}[t!]
\includegraphics[width=0.5\textwidth]{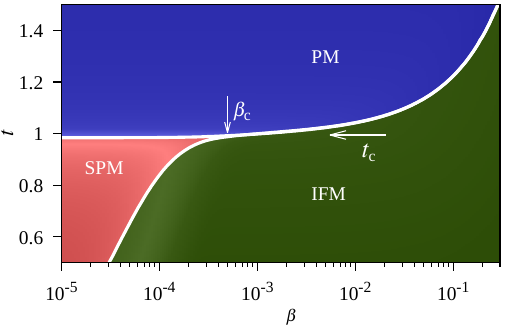}
\caption{Phase diagram in coordinates ($t, \beta$). The diagram shows paramagnetic (PM), superparamagnetic (SMP)  and induced ferromagnetic (IFM) phases.
The arrows show a magnetic field $\beta_c$ and a temperature $t_c$ which correspond to the absolute maximum of $\chi_2$, $N=16384, na^3=0.03$.
}
\label{phase}
\end{figure}

Assuming that the phase transition correspond to the maxima of magnetic susceptibility $\chi$ and $\chi_2$, 
one can plot the phase diagram shown in figure \ref{phase}.
It should be noted that the positions of the maxima of the magnetic susceptibilities $\chi$ and $\chi_2$ do not coincide.
The maximum of $\chi_2$ is associated with the parallel orientation of individual spins inside a cluster.
At temperatures higher than this maximum the spin system is paramagnetic (PM).
The maximum of the magnetic susceptibility $\chi$ is associated with the orientation of the magnetic moment of the whole cluster by the magnetic field.
In weak magnetic fields only the average square of the magnetic moment changes, while the average magnetic
moment remains close to zero \cite{demishev2020electron, bogoslovskiy2021spin, demishev2022spin} and the spin system is in a superparamagnetic state (SPM).
This is a consequence of the fact that we consider a system with a large but finite number of spins $N$
whose properties are governed by thermodynamic fluctuations. If the magnetic field is high and the temperature is low
spins are oriented along the magnetic field and the system is in a ferromagnetic state induced by the magnetic field (IFM).

\begin{figure}[b]
\includegraphics[width=0.5\textwidth]{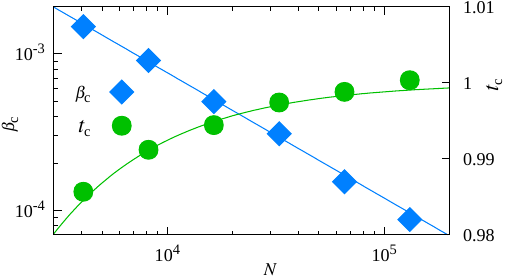}
\caption{Dependencies of $\beta_c$ and $t_c$ on the number of spins in the system $N$. Lines are empirical fits with functions \mbox{$\beta_c = 5/4 N^{-4/5}, t_c = -10\beta_c + 1$}.
}
\label{b_c_t_c}
\end{figure}

Phase diagram is dependent on the number of spins. In the paper we show color plots which are calculated for $N=16384$.
In supplementary materials we include all color plots for different $N$.
Results of calculations for different $N$ values are summarized at figure \ref{b_c_t_c}.
Our calculations show that $\beta_c$ and $t_c$  depend on $N$ in accordance with a~power law.

\begin{figure}[!t]
\centering
\includegraphics[width=1\linewidth]{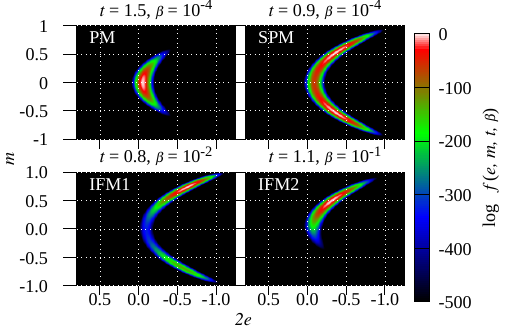}
\caption{Logarithm of the probability density function $f (e, m, t, \beta)$ for different values of temperature and magnetic field. }
\label{f}
\end{figure}

The dependence on $N$ can be understood from (\ref{Z0}) and (\ref{mean_M}).
The two terms in (\ref{mean_M}) $m_+Z_+$ and $m_-Z_-$ exponentially depend on $N$.
This means that the ratio between them is strongly dependent on the size of the system.
At small $N$, two terms are comparable, thermodynamic fluctuations are large, and the cluster is in the superparamagnetic state.
As $N$ increases, the number of states with a magnetic moment directed along the magnetic field becomes much larger,
and the cluster becomes ferromagnetic. This behavior is illustrated on figure \ref{f} where the probability density
function (\ref{probability}) in diffirent phases is plotted. In the case of an infinite system the probability
maxima are just $\delta$-functions. If $N$ is finite, the system has a non-zero probability to be
in a number of states around $m_0$ or $m_+$ and $m_-$ maxima and can switch between them.

The critical field $\beta_c$, above which the superparamagnetic phase does not exist,
decreases with increasing $N$ according to a power law (see figure \ref{b_c_t_c}).
In a semiconductor with magnetic impurities,
many oriented magnetic moments of individual clusters create a Weiss molecular field.
As the concentration of magnetic impurities increases, both the average number of spins in these clusters
and the Weiss field increase.
At some point, the value of this field exceeds the value of the critical field
$\beta_c$ and the semiconductor passes into the ferromagnetic
state.

In the paramagnetic state the probability density function has only one maximum at the magnetic moment which is close to zero (Fig.~\ref{f}).
When temperature is lower than $t_c$ the probability density function has two
close maxima, the system is superparamagnetic and switches between these maxima due to thermodynamic fluctuations.
At $\beta > \beta_c$ one of the probability maxima is much stronger then the other and the system is in the IFM1 state.
If the magnetic field is high enough, there is only one probability maximum which we mark as IFM2 state.

We also would like to note that from (\ref{mean_E}) the average exchange energy linearly depends
on the spin concentration for an arbitrary $J(r)$.
Therefore, the model predicts that the Curie temperature $T_c(n) = \overline{J_1}$ linearly depends on the spin concentration, 
at least in the region where $J(r)$ is independent of $n$.

In conclusion, an analytical model is developed that describes a system of a finite number of randomly distributed spins, 
taking into account the long-range Ising-type exchange interaction and the magnetic field.
The phase diagram in coordinates ($t, \beta$) explains the complex nature of the phase transition in ferromagnetic semiconductors
and qualitatively agrees with experimental results.
The origin of the intermediate superparamagnetic phase and its relation to thermodynamic fluctuations
in finite-size systems are explained.

\section{Acknowledgement}

We acknowledge support from Russian Science Foundation (Grant No. 23-22-00333)
The numerical calculation were performed using computational resources of the supercomputer center in Peter the Great Saint-Petersburg Polytechnic University Supercomputing Center.


\section{Appendix}

Similarly with the mean exchange energy $\overline E_m$, we can calculate the variance of the exchange energy
\begin{equation} 
\nonumber
\sigma^2 = \overline {E^2} - {\overline {E}}^2 
\end{equation}
We use the explicit expression for the exchange energy (\ref{exchange_energy})
\begin{equation} 
\nonumber
\sigma^2 = \frac{1}{4} \overline {\sum \limits_{\substack{i, j \\ k, m}} J_{ij} J_{k l} s_i s_j s_k s_l} - 
\frac{1}{2} \overline {\sum \limits_{i, j} J_{i j}  s_i s_j } \frac{1}{2} \overline {\sum \limits_{k, l} J_{kl} s_k s_l}
\end{equation}

In this paper we consider a system of randomly oriented spins. This means that there is no correlation between the value of the exchange energy $ J_{ij}$ and the direction of the spin $s_i$. 
Therefore, averaging over coordinates and over spin directions could be carried out separately.
\begin{equation} 
\sigma^2 = \frac{1}{4} \overline {\sum \limits_{\substack{i, j \\ k, l}} J_{ij} J_{kl}} \overline { s_i s_j s_k s_l} - 
\frac{1}{2} \overline {\sum \limits_{i, j} J_{ij}}  \overline {s_i s_j }  \frac{1}{2} \overline {\sum \limits_{k, l} J_{kl}} \overline { s_k s_l}
\label{sigma2}
\end{equation}

In this equation we will separately consider the terms for which all 4 indices $(i, j, k, l)$ are different, two indices coincide and two pairs of indices coincide.
First we consieder the case when all indices are different. Averaging over a pair of spins in an explicit form gives
\begin{align*} 
\nonumber
\overline {s_i s_j } = \frac{q(q-1) + (N-q)(N-q-1) - 2q(N-q)}{N(N-1)} = \\
m^2+ \frac{m^2 -1}{N} + O\left( \frac{1}{N^2} \right)
\end{align*}

Averaging over four spin variables with different indices gives
\begin{equation} 
\nonumber
\overline { s_i s_j s_k s_l} = m^4 - \frac {6m^2}{N} + \frac {6m^4}{N} + O\left( \frac{1}{N^2} \right)
\end{equation}

Averaging over coordinates gives \ref{mean_E}
\begin{equation} 
\nonumber
\frac{1}{2} \overline {\sum \limits_{i, j} J_{ij}} = \overline{J} = \frac{1}{2}N\overline{J_1}
\end{equation}

In the sum over $k,l$ in (\ref{sigma2}), the indices $k, l$ must not coincide with the indices $i,j$ from the first sum. Then for each of the indices $k, l$ there are only $N-2$ possible values
\begin{equation} 
\nonumber
\frac{1}{2} \overline {\sum \limits_{k, l} J_{kl}} = \frac{1}{2}\frac{(N - 2)^2}{N}\overline{J_1} = \frac{1}{2} \left( N - 4 \right) \overline{J_1} + O\left( \frac{1}{N} \right)
\end{equation}

The sum $\sum \limits_{i, j, k, l} J_{ij} J_{kl} $ can be considered as the product of two sums, $i, j$ and $k, l$. Since we are considering the case where all indices are different, averaging in these sums can be carried out independently
\begin{equation} 
\nonumber
\frac{1}{4} \overline {\sum \limits_{\substack{i, j \\ k, l}} J_{ij} J_{kl}} = \frac{1}{4} \left( N^2- 4N \right) \overline{J_1}^2 + O\left( 1 \right)
\end{equation}

We substitute these expressions into the variance (\ref{sigma2}). The terms of order $N^2$ cancel and we obtain that the terms with 4 different indices in (\ref{sigma2}) give
\begin{equation}
N \overline{J_1}^2 (m^4-m^2) + O\left( 1 \right)
\label{4index}
\end{equation}

Next we consider the terms with exactly two matching indices in the expression (\ref{sigma2}). We will denote the matching indices by $i$, and different indices by $j$ and $k$. 
There are 4 possible options for equal indices in the original notation $(i = k, i = l, j = k, j = l)$, therefore after redesignation the multipliers $\frac{1}{4}$ and $\frac{1}{2}$ in (\ref{sigma2}) will be cancelled. 
In the second term of the expression, we rewrite the two sums over $i, j$ and $i, k$ as a total sum over three indices.
\begin{align} 
\label{ijk}
\overline {\sum \limits_{\substack{i, j, k \\ k \neq j}} J_{ij} J_{ik}} \overline {s_j s_k} - 
\overline {\sum \limits_{\substack{i, j, k \\ k \neq j}} J_{ij} J_{ik}} \overline {s_i s_j } \; \overline { s_i s_k} = \\
N \overline{J_1}^2 (m^2 - m^4) + O\left( 1 \right)
\nonumber
\end{align}
And the terms with two pairs of equal indices give
\begin{equation}
\nonumber
\frac{1}{2} \overline {\sum \limits_{i, j} J_{ij}^2} - \frac{1}{2} \overline {\sum \limits_{i, j} J_{ij}^2}  \overline {s_i s_j } \; \overline {s_i s_j } = 
\frac{1}{2} \overline {\sum \limits_{i, j} J_{ij}^2} \left( 1 - m^4 \right)
\end{equation}

Note that the expression (\ref{ijk}) coincides, up to the minus sign, with the contribution from terms with 4 different indices (\ref{4index}).
These terms cancel each other out. Finally we obtain the following expression for the variance
\begin{equation} 
\nonumber
\sigma^2 = 
\frac{1}{2} \overline {\sum \limits_{i, j} J_{ij}^2} \left( 1 - m^4 \right) = 
\sigma_0^2 \left( 1 - m^4 \right)
\end{equation}

Here we introduce the notation $\sigma_0^2$, which is equal to the variance of the exchange energy at zero magnetic moment. 
In order to calculate $\sigma_0^2$, we first calculate the square of the exchange energy for two spins with numbers $i$ and $j$, the distance between which is no more than $R$

\begin{equation} 
\nonumber
\overline {J^2_{ij}(R)} = \frac{3}{4 \pi R^3} \int \limits_0^R J^2(r) 4 \pi r^2 dr
\end{equation}

We substitute the explicit expression for $J(r)$ and tend the upper limit to infinity.

\begin{equation} 
\nonumber
\overline {J^2_{ij}} = \frac{3 J_0^2 a^3}{R^3 2^{16}} \int \limits_0^{\infty} \left(\frac{4r}{a} \right)^{7} \exp \left( - \frac{4 r}{a} \right) \,d {\left(\frac{4 r}{a} \right)}
\end{equation}

After integration we get
\begin{equation} 
\nonumber
\overline {J^2_{i,j}} = \frac{3\, J_0^2 a^3\, 7!}{2^{16} R^3}
\end{equation}

Since in our model there is no correlation in the arrangement of spins, the sum of the squares of the exchange energies of spin number $i$ with all spins from a ball of radius $R$ will be equal to
\begin{equation} 
\nonumber
\overline {\sum \limits_{j} J^2_{ij}} = n \frac{4}{3}\pi R^3 \overline{J^2_{ij}} = \frac{7!\, \pi J_0^2 n a^3}{2^{14}}
\end{equation}

Finally, we sum over all spins $i$ and write the multiplier $\frac{1}{2}$ because each exchange energy is included in the sum 2 times.
\begin{equation} 
\nonumber
\sigma_0^2 = \frac{1}{2} \overline {\sum \limits_{i,j} J^2_{ij}} = N \frac{7!}{2^{15}} \pi J_0^2 n a^3
\end{equation}

\vspace{3 cm}



%

\end{document}